\documentclass[10pt]{amsart}
\usepackage{amssymb}
\usepackage{enumerate}
\usepackage[dvips]{graphicx}
\DeclareGraphicsExtensions{.eps,.art,.ART,.ps}
\newcommand{\bbR}{\mathbb{R}}      
\newcommand{\bbN}{\mathbb{N}}      
\newcommand{\link}{\operatorname{link}}     
\newcommand{\wind}{\operatorname{wind}}     
\newtheorem{Theo}{Theorem}[section]
\newtheorem{Prop}[Theo]{Proposition}
\newtheorem{Lemma}[Theo]{Lemma}

\newtheorem{Cor}[Theo]{Corollary}
\newtheorem{Con}[Theo]{Conjecture}
\theoremstyle{definition}
\newtheorem{Def}[Theo]{Definition}
\theoremstyle{remark}

\begin{document}
\begin{abstract}
The set $N$ of all null geodesics of a globally hyperbolic $(d+1)$-dimensional spacetime $(M,g)$ is naturally a smooth $(2d-1)$-dimensional contact manifold. The \emph{sky} of an event $x \in M$ is the subset
\[
X = \left\{ \gamma \in N:x \in \gamma \right\} \subset N
\]
and is an embedded Legendrian submanifold of $N$ diffeomorphic to $S^{d-1}$. It was conjectured by Low that for $d=2$ events $x,y \in M$ are causally related \emph{iff} $X,Y \subset N$ are linked (in an appropriate sense). We use the contact structure and knot polynomial calculations to prove this conjecture in certain particular cases, and suggest that for $d=3$ smooth linking should be replaced with Legendrian linking.
\end{abstract}
%
%
\title{Linking, Legendrian linking and causality}
\author{Jos\'{e} Nat\'{a}rio}
\address{Mathematical Institute, Oxford}
\curraddr{Department of Mathematics, Instituto Superior T\'{e}cnico, Portugal}
\author{Paul Tod}
\address{Mathematical Institute, Oxford}
\thanks{The first author was partially supported by FCT (Portugal) through programs PRAXIS XXI and POCTI}
\subjclass[2000]{Primary 83C60; Secondary 53D10, 57M25.}
\maketitle
%
%
%
\section{Introduction} \label{section1}
The set $N$ of all null geodesics of a given spacetime $(M,g)$ plays an important role in twistor theory (see \cite{PR86}). If $(M,g)$ is a globally hyperbolic $(d+1)$-dimensional spacetime then $N$ is naturally a smooth $(2d-1)$-dimensional contact manifold. The \emph{sky} of an event $x \in M$ is the subset
\[
X = \left\{ \gamma \in N:x \in \gamma \right\} \subset N
\]
and is an embedded Legendrian submanifold of $N$ diffeomorphic to $S^{d-1}$.

It is natural to ask how the geometry of events in $M$ is related to the geometry of the corresponding skies in $N$. In particular, it was conjectured by Low (see \cite{L88}) that for $d=2$ events $x,y \in M$ are causally related \emph{iff} $X,Y \subset N$ are linked (in an appropriate sense). We will use the contact structure of $N$ and knot polynomial calculations to prove this conjecture in certain particular cases. We will also address the $d=3$ case, where we conjecture that smooth linking should be replaced with Legendrian linking.

This paper is organized as follows. In section \ref{section2} we describe the differential and contact structures of the manifold of light rays $N$ of a general globally hyperbolic $(d+1)$-dimensional spacetime $(M,g)$. It turns out that if $\Sigma \subset M$ is any Cauchy surface endowed with the Riemannian metric induced by $g$, $N$ can be naturally identified with the cotangent sphere bundle $TS^* \Sigma$; the contact structure is then seen to be the one arising from the natural symplectic structure of $T^* \Sigma$. The sky $X \subset N$ of a spacetime point $x \in M$ is defined and shown to be a Legendrian submanifold of $N$. It is seen that a sky $X\in N$ can be obtained from its (cooriented) wavefront $\pi(X) \subset \Sigma$, where $\pi:TS^*\Sigma \to \Sigma$ is the natural projection. The section ends with a review of the notions of linking and Legendrian linking.

In section \ref{section3} we turn to the $d=2$ case. We show how to compute the linking number of two skies from the corresponding wavefronts and analyze a few simple examples in order to motivate Low's conjecture, which we then state. Much of the material in the first two sections was first discussed by Low (see \cite{L88}, \cite{L89} \cite{L90a}, \cite{L90b}, \cite{L94}, \cite{L98}), and can be found in greater detail in \cite{N00}.

In section \ref{section4} we use the fact that skies $X \subset N$ are Legendrian submanifolds to determine the possible configurations of wavefronts $\pi(X) \subset \Sigma$. This provides a heuristic explanation of why Low's conjecture should be true, which we make rigorous in section \ref{section5} for a large class of examples by computing certain coefficients of their Kauffman polynomials.

Finally we consider the $d=3$ case in section \ref{section6}. We give an example of skies of causally related points which are unlinked, and conjecture that smooth linking should be replaced with Legendrian linking in the correct version of Low's conjecture for this dimension.
%
%
%
\section{The manifold of light rays} \label{section2}
Let $(M,g)$ be a $(d+1)$-dimensional globally hyperbolic spacetime with a fixed time orientation, and consider the set $N$ of all its null geodesics (we define a null vector to be a \emph{nonzero} vector $v$ such that $g(v,v)=0$, and hence the constant geodesic is \emph{not} a null geodesic). If $\Sigma \subset M$ is a Cauchy surface, then every null geodesic intersects $\Sigma$ exactly once. On the other hand, at any event $x \in \Sigma$ two future-pointing null vectors are initial conditions for the same null geodesic \emph{iff} they are linearly dependent, or, equivalently, \emph{iff} their orthogonal projections on $T_x\Sigma$ are linearly dependent and point in the same direction. Consequently, $N$ can be identified with the tangent sphere bundle $TS\Sigma$ (recall that $\Sigma$ endowed with the metric induced by $g$ is a Riemannian manifold of dimension $d$). We use this fact to define a differentiable structure on $N$, and notice that, due to smooth dependence of the solutions of the geodesic equation on its initial data, this structure is independent of the choice of Cauchy surface. Thus the set of all null geodesics of a globally hyperbolic $(d+1)$-dimensional spacetime is a differentiable manifold of dimension $d+(d-1)=2d-1$, which we call its \emph{manifold of light rays}.

It is possible to show that $N$ has a natural contact structure. In order to do so, we introduce the so-called {\em manifold of scaled light rays} $\tilde{N}$. This manifold is defined analogously to the manifold of light rays, except that we distinguish between null geodesics with different affine parameterizations; more precisely, if $\gamma_1, \gamma_2 : \bbR \to M$ are two affinely parametrized null geodesics such that $\frac{d}{ds_1}{\gamma_1}, \frac{d}{ds_2}{\gamma_2}$ are future pointing, we take $\gamma_1 = \gamma_2$ {\em iff} $\gamma_1(\bbR)=\gamma_2(\bbR)$ and $\frac{d}{ds_1}\left(\gamma_2^{-1} \circ \gamma_1\right) = 1$. Therefore $\tilde{N}$ can be identified with the tangent bundle to a Cauchy surface minus the zero section. In particular, $\dim\tilde{N}=2d$. A smooth path in $\tilde{N}$ can always be represented by a smooth function $\gamma: \bbR^2 \to M$ such that $\gamma(s, \alpha)$ is, for each value of $\alpha$, an affinely parametrized null geodesic. If we define the vector fields \footnote{We shall use the notation conventions of \cite{PR86}, where latin indices represent abstract indices.}
\begin{align*}
p^a=\gamma_* \frac{\partial}{\partial s} \\
X^a=\gamma_* \frac{\partial}{\partial \alpha}
\end{align*}
and the operator
\[
D=p^a \nabla_a
\]
it is a simple matter to obtain the identities
\begin{align*}
&Dp^a=0 \,\,\,\, \text{(geodesic equation)};\\
&\left[\frac{\partial}{\partial s},\frac{\partial}{\partial \alpha}\right]=0 \Rightarrow  DX^a=X^b\nabla_bp^a;\\
&X^b\nabla_b\left(p_ap^a\right)= 0 \Rightarrow p_aX^b\nabla_b p^a=0 \Rightarrow p_a DX^a = 0;\\
&D^2 X^a=R_{bcd}^{\,\,\,\,\,\,\,\,a}p^bX^cp^d \,\,\,\, \text{(i.e., } X^a\text{ is a Jacobi field)},
\end{align*}
where $R_{bcd}^{\,\,\,\,\,\,\,\,a}$ is the Riemann tensor. Thus $\gamma(s, \alpha)$ yields a Jacobi field along the geodesic $\gamma=\gamma(\cdot,0)$ satisfying $p_a DX^a = 0$. Such fields form a vector space $J_\gamma$ of dimension $2(d+1)-1=2d+1$; thus the natural linear surjective map $\Pi: J_\gamma \to T_\gamma \tilde{N}$ has kernel of dimension $1$, which is easily seen to be the set $I_\gamma$ of vector fields of the form $X^a=kp^a$ for $k \in \bbR$ (as these correspond to going to the same affinely parametrized geodesics). In other words,
\[
T_\gamma \tilde{N}=\frac{J_\gamma}{I_\gamma}.
\]
If $X^a, Y^b \in J_\gamma$, then
\[
D(p_aX^a)=p_a DX^a=0
\]
and
\[
D(Y_a DX^a - X_a DY^a) = Y_a R_{bcd}^{\,\,\,\,\,\,\,\,a}p^bX^cp^d - X_a R_{bcd}^{\,\,\,\,\,\,\,\,a}p^bY^cp^d = 0.
\]
(the last equality, known as {\em Lagrange's identity}, following from the symmetries of the Riemann tensor). Thus we can use the above expressions to define a 1-covector $\Theta$ and a 2-covector $\Omega$ on $J_\gamma$. If $X^a \in I_\gamma$, then  $X^a=kp^a$ and
\[
p_aX^a=kp_ap^a=0
\]
and
\[
Y_a DX^a - X_a DY_a = kY_a Dp^a- k p_a DY^a=0.
\]
Consequently there exist a 1-covector $\tilde{\theta}$ and a 2-covector $\tilde{\omega}$ on $T_\gamma \tilde{N}$ such that $\Theta = \Pi^* \tilde{\theta}$, $\Omega=\Pi^*\tilde{\omega}$. Using local coordinates, it is not hard to show that $\tilde{\theta}=i^*\theta,\tilde{\omega}=i^*\omega$, where $i:T\Sigma \to T^*\Sigma$ is minus the natural map determined by the Riemannian metric $h$ induced on $\Sigma$ by the spacetime metric $g$, and $\theta,\omega$ are the canonical symplectic forms on $T^*\Sigma$ (the minus sign arises from our spacetime signature convention, which is $(+,-,-,-)$). In other words, the manifold of scaled light rays is naturally a symplectic manifold; the symplectic structure is just the natural symplectic structure of $T^*\Sigma$, provided that we identify $\tilde{N}\simeq T\Sigma \setminus 0\Sigma$ with $T^*\Sigma \setminus 0^*\Sigma$ using the map $i$ ($0\Sigma$ and $0^*\Sigma$ being the zero sections of $T\Sigma$ and $T^*\Sigma$). Notice that although we are giving the symplectic forms in terms of a particular Cauchy surface $\Sigma$, the symplectic structure of $\tilde{N}$ is (by construction) clearly independent of the choice of $\Sigma$.

To obtain the manifold of light rays $N$ from the manifold of scaled light rays $\tilde{N}$ we have to identify geodesics which differ by an affine reparameterization. This corresponds to quotienting $\tilde{N}\simeq T\Sigma \setminus 0\Sigma\simeq T^*\Sigma \setminus 0^*\Sigma$ by the natural action of $\bbR^+$ along the fibers. Hence $N$ has a natural contact structure; we can think of $N$ as the cotangent sphere bundle $T^*S\Sigma \subset T^*\Sigma \setminus 0^*\Sigma$ (taken with respect to an arbitrary Riemannian metric on $\Sigma$, for instance $h$), in which case a contact 1-form is simply the restriction of $\theta$ to $N$ (see \cite{Ar97}).

The \emph{sky} of an event $x \in M$ is the subset
\[
X = \left\{ \gamma \in N:x \in \gamma \right\} \subset N.
\]

If $x \in \Sigma$ then $X$ is a fiber of the cotangent sphere bundle $T^*S\Sigma$, and therefore is an embedded submanifold of $N$ diffeomorphic to $S^{d-1}$. Since we are free to regard $N$ as a fiber bundle over \emph{any} Cauchy surface, and every event in $M$ belongs to \emph{some} Cauchy surface, we see that any sky is an embedded $S^{d-1}$.

Furthermore, since the contact form $\theta$ vanishes on the fibers $T^*S\Sigma$, we see that the sky of any point is a Legendrian submanifold of $N$ (i.e., a maximal dimension submanifold of $N$ where the contact form vanishes). Notice that in particular this means that the null cone of any event $x\in M$ is orthogonal to the tangent vectors of the generating null geodesics (i.e., the null cone is a null hypersurface).

The {\em wavefront} generated by the event $x \in M$ at $\Sigma$ is simply the intersection of its null cone with $\Sigma$ (or, equivalently, $\pi(X)$, where $\pi:N \to \Sigma$ is the natural projection). Given a wavefront, there exist at each point of the wavefront two future-pointing null directions orthogonal to the wavefront, whose orthogonal projections on $T\Sigma$ are negative multiples of each other. Thus we can reconstruct $X$ from the wavefront $\pi(X)$ provided that we are given the {\em coorientation} of the wavefront, i.e., a choice between the two possible null directions at each point. This extra bit of information can be given, for instance, as a unit normal vector field ${\bf n}$ on $\pi(X)$ (with respect to the Riemannian metric $h$, say). Notice that ${\bf n}$ is necessarily continuous, and hence it suffices to indicate it at any single point of $\pi(X)$.

Let us now assume that $M$ is orientable. Then by choosing a global time function $t:M \to \bbR$ and using the globally defined nonvanishing future-pointing timelike vector field $\nabla_a t$ we can orient all Cauchy surfaces $\{t=\text{constant}\}$; hence we can define an orientation on each sky by orienting tangent spheres on each Cauchy surface (it is easy to check that this orientation does not depend on the choice of global time function).

If $x,y \in M$ are not in the same null geodesic then $X \sqcup Y$ is a \emph{smooth link}, i.e., a disjoint union of embedded $S^{d-1}$s in a smooth manifold of dimension $2d-1$. Recall that a smooth one-parameter family of diffeomorphisms $\Phi_{t}:[0,1] \times N \to N$ is said to be a \emph{smooth (ambient) isotopy} if $\Phi_{0}$ is the identity map. Two links $X_0 \sqcup Y_0$ and $X_1 \sqcup Y_1$ are said to be \emph{equivalent} if there exists a smooth isotopy $\Phi_{t}:[0,1] \times N \to N$  such that $\Phi_{1}(X_0)=X_1$ and  $\Phi_{1}(Y_0)=Y_1$. It can be shown that two links $X_0 \sqcup Y_0$ and $X_1 \sqcup Y_1$ are equivalent \emph{iff} there exists a \emph{smooth motion} carrying one into the other, i.e., \emph{iff} there exist smooth one-parameter families of embeddings $f_{t}:S^{d-1} \to N$ and $g_{t}:S^{d-1} \to N$ such that $f_0(S^{d-1})=X_0$, $g_0(S^{d-1})=Y_0$, $f_1(S^{d-1})=X_1$, $g_1(S^{d-1})=Y_1$ and $f_t(S^{d-1}) \cap g_t(S^{d-1}) = \varnothing$ for all $t \in [0,1]$.

If $(x_0, y_0)$ and $(x_1, y_1)$ are two pairs of non-causally related events, one can easily construct smooth curves $\alpha,\beta:[0,1] \to M$ such that $\alpha(0)=x_0$,  $\beta(0)=y_0$, $\alpha(1)=x_1$, $\beta(1)=y_1$ and $\alpha(t), \beta(t)$ are not in the same null geodesic for all $t \in [0,1]$. This induces a smooth motion of $X_0 \sqcup Y_0$ into  $X_1 \sqcup Y_1$, which are therefore equivalent. We conclude that the link formed by the skies of any two non-causally related events belong to the same equivalence class, which we define as the \emph{unlink} in $N$ (notice that although there exists a natural choice for the unlink in  $\bbR^{2d-1}$ or $S^{2d-1}$, this is not the case in general for an arbitrary smooth $(2d-1)$-dimensional manifold). $X,Y \subset N$ are said to be {\em linked} if the equivalence class of $X \sqcup Y$ is not the unlink.

A {\em Legendrian link} is a smooth link $X \sqcup Y$ such that $X,Y \subset N$ are Legendrian. Recall that a diffeomorphism $\Phi: N \to N$ is said to be a {\em contactomorphism} if it preserves the contact structure, i.e., if $\Phi_* \ker \theta = \ker \theta$. In particular, a contactomorphism maps Legendrian submanifolds to Legendrian submanifolds. A {\em Legendrian isotopy} is a smooth isotopy in which each diffeomorphism is a contactomorphism. A {\em Legendrian embedding} $f:S^{d-1} \to N$ is a smooth embedding whose image is Legendrian. All that was said above about smooth links remains true if we replace ``smooth'' with ``Legendrian''. Thus if $x,y \in M$ are not in the same null geodesic then $X \sqcup Y$ is a Legendrian link, and if $(x_0, y_0)$ and $(x_1, y_1)$ are two pairs of non-causally related events, then $X_0 \sqcup Y_0$ and $X_1 \sqcup Y_1$ are Legendrian equivalent (and we define the corresponding equivalence class as the {\em Legendrian} unlink).

Notice that if  $X_0 \sqcup Y_0$ and $X_1 \sqcup Y_1$ are Legendrian equivalent, they must be (smooth) equivalent, but the reverse is not true in general. We shall present an example of this later on.
%
%
%
\section{Low's conjecture} \label{section3}
Let us consider now the case where $d=2$ and $\Sigma$ is diffeomorphic to a (connected, open) subset of $\bbR^2$. In this case, $N$ is diffeomorphic to a subset of the tangent sphere bundle of $\bbR^2$, which in turn is diffeomorphic to the interior of a solid torus in $\bbR^3$. It will prove useful to fix a particular embedding $\sigma:TS\bbR^2 \to \bbR^3$ (which we shall call the {\em standard embedding}). Thus, if $(r,\theta)$ are the usual polar coordinates in $\bbR^2$, and $\varphi$ is the coordinate in the fibers of $TS\bbR^2$ corresponding to the angle with the positive $x^1$-direction, we define
\[
\sigma(r,\theta,\phi) = ((2+\tanh r \cos\theta)\cos\varphi,(2+\tanh r \cos\theta)\sin\varphi, \tanh r \sin \theta)
\]

This particular embedding has the advantage of carrying the unlink in $N$ (as defined above) to the unlink in $\bbR^3$. Notice that the skies of events on $\Sigma$ are mapped to circles of constant $(r,\theta)$, which for convenience we call {\em meridians}. From now on we shall identify  $N$ with $\sigma \left( N \right)$. Clearly a smooth motion of a smooth link in $N$ yields a smooth motion of this link in $\bbR^3$, and consequently if two links are equivalent in $N$ they must be equivalent in $\bbR^3$.

Recall that in $\bbR^3$ there is a well-known link invariant, namely the {\em linking number}. Given a link $X \sqcup Y$, this is simply the integer given by the {\em Gauss integral},

\[
\link(X,Y)=\frac{1}{4 \pi}\oint_{X} \oint_{Y} \frac{({\bf r}-{\bf s}) \cdot (d{\bf r}\times d{\bf s})}{\left\| {\bf r}-{\bf s} \right\|^3}
\]

\noindent (where ${\bf r},{\bf s}:S^1 \to \bbR^3$ are embeddings for $X,Y$), or, equivalently, the integer such that $\left[ X \right]$ is $\link(X,Y)$ times an appropriate generator of $H_1(\bbR^3 \setminus Y)$, depending on the orientation of $Y$ (see \cite{R90}). In particular if  $X \sqcup Y$ is the unlink then $\link(X,Y)=0$. Notice also that, as is easily seen from the Gauss integral formula, $\link(X,Y)=\link(Y,X)$.

As was noted by Low (\cite{L88}), $\link(X,Y)$ is simply the winding number of the wavefront generated by $x$ at a Cauchy surface through $y$ around $y$. Indeed, let $x,y \in M$ be two events not in the same null geodesic. Recall that we can always assume that $y \in \Sigma$, i.e., that $Y$ is a meridian. In general, however, $x \not\in \Sigma$. We can think of the natural projection $\pi: N \to \Sigma$ as quotienting $N$ by its meridians, and consequently identify $\Sigma$ with the quotient space. Thus any smooth surface on $N$ intersecting each meridian exactly once can be identified with $\Sigma$. In this manner we can think of $y$ and $\pi(X)$ as a point and an oriented curve on $\Sigma \subset N$ (since $X$ is an oriented curve, $\pi(X)$ can be given the induced orientation).

One can always choose $\Sigma \subset N$ such that $\Sigma \cap X$ is finite. It is then possible to construct an isotopy of $N$ deforming $X$ into a curve which approaches (with any required accuracy) the wavefront $\pi(X)$ plus a finite number of meridians, one hanging from each intersection of $X$ and $\Sigma$. Indeed, if $\varphi$ is the angular coordinate along the meridians such that $\Sigma=\{\varphi=0\}$, one has but to consider isotopies of the form $\varphi \mapsto \varphi + \delta(t, \varphi)$, where $\delta:[0,1] \times [0,2 \pi] \to [0, 2 \pi]$ is a nonnegative smooth function vanishing for $t=0$ and $\varphi=0, 2 \pi$ and approaching $2 \pi - \varphi$ from below as $t \to 1$. Since the linking number of any two disjoint meridians is zero, we see that $\link(X,Y)=\wind(\pi(X),y)$, where $\wind(\pi(X),y)$ is the winding number of the curve $\pi(X) \subset \Sigma$ around the point $y \in \Sigma$ (recall that $\Sigma$ is diffeomorphic to a subset of $\bbR^2$).

(This result plus the fact that $\link(X,Y)=\link(Y,X)$ allows us to make the following nontrivial observation: if $x,y$ are not in the same null geodesic and $\Sigma_x, \Sigma_y$ are arbitrary Cauchy surfaces through $x,y$ then the winding number of the wavefront generated by $x$ at $\Sigma_y$ around $y$ is equal to the winding number of the wavefront generated by $y$ at $\Sigma_x$ around $x$).

As an example, consider Minkowski's $(2+1)$-dimensional spacetime, i.e., $\bbR^3$ endowed with the line element $ds^2=dt^2-\left(dx^1\right)^2-\left(dx^2\right)^2$, and take hypersurfaces of constant $t$ as Cauchy surfaces. Then all wavefronts are circles, and all linking numbers are therefore either zero or one. Since an event on the Cauchy surface is causally related to the event generating the wavefront {\em iff} it is either inside the wavefront (in which case the winding number is $1$) or on it, we see that two events in Minkowski $(2+1)$-dimensional spacetime are causally related {\em iff} their skies either intersect or are linked with linking number $1$.

\begin{figure}[h]
\begin{center}    
        \includegraphics[scale=.4]{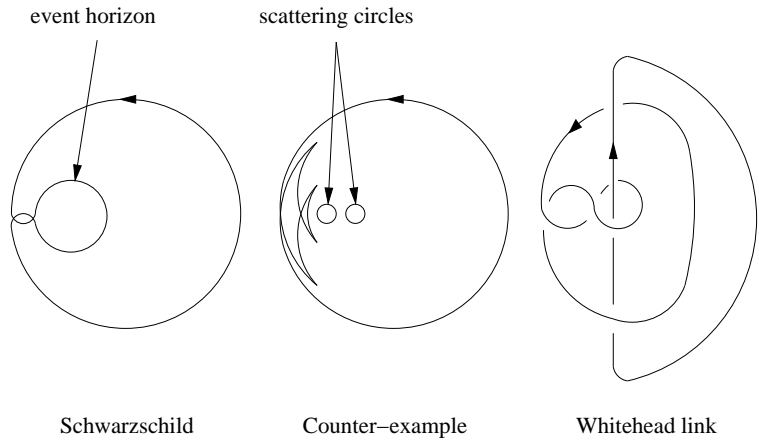}
\end{center}
\caption{Examples of wavefronts and an associated link.}
\label{picture}
\end{figure}

Another example is provided by Schwarzschild's $(2+1)$-dimensional static spacetime, i.e., the region of $\bbR^3$ given in cylindrical coordinates $(t,r,\varphi)$ by $r>1$ endowed with the line element $ds^2=\left(1-\frac{1}{r}\right)dt^2-\left(1-\frac{1}{r}\right)^{-1}dr^2-r^2d\varphi^2$ (we've taken the Schwarzschild radius as our length unit). If we again take hypersurfaces of constant $t$ as Cauchy surfaces, it is possible to show that the wavefronts are as shown in figure \ref{picture}, wrapping around the event horizon any number of times. It is then easily seen that all winding numbers are either zero or positive, and that an event in the Cauchy surface not on the wavefront is causally related to the event generating the wavefront {\em iff} the winding number is positive. Consequently, two events in Schwarzschild $(2+1)$-dimensional static spacetime are causally related {\em iff} their skies either intersect or are linked with positive linking number (but links do occur with any positive linking number).

It is not true in general that all links formed by the skies of causally related events have nonvanishing linking number. A simple counter-example is provided by $\bbR^3$ endowed with the line element $ds^2=dt^2-\Omega^2\left(x^1,x^2\right)\left(\left(dx^1\right)^2+\left(dx^2\right)^2\right)$, where $\Omega:\bbR^2 \to [1, + \infty)$ is an appropriate smooth function (equal to $1$ except on the circles $\left(x^1\right)^2+\left(x^2\right)^2<1$ or $\left(x^1+4\right)^2+\left(x^2\right)^2<1$, where it increases radially towards the center). The wavefront of the event $(0,4,0)$ on the Cauchy surface $t=10$ is as depicted in figure \ref{picture}, each pair of cusps corresponding to scattering by one of the circles where the metric is not flat. The appearance of the second pair of cusps allows the existence of events $y$ that, although clearly causally related to $x$, are such that the winding number of $\pi(X)$ around $y$ is zero, and consequently $\link(X,Y)=0$. However $X \sqcup Y$ is {\em not} the unlink. To see that, we notice that the Riemannian metric induced on $\Sigma$ is conformally related to the Euclidean metric, and that we can therefore use the usual angle with the $x$-axis as a coordinate on $TS\Sigma$. To decide on the value of this coordinate along the wavefront, we recall that the tangent vector $\dot{t} \frac{\partial}{\partial t}+\dot{x}^1\frac{\partial}{\partial x^1}+\dot{x}^2\frac{\partial}{\partial x^2}$ to any null geodesic is orthogonal to the light cone of $x$. Consequently, the element of $TS\Sigma$ corresponding to the null geodesic is the normal vector (on $\Sigma$) to the wavefront, oriented so that it points outwards when the wavefront is in the boundary of the causal future. Using these rules one constructs the link shown in figure \ref{picture}, which is the so-called {\em Whitehead link} (and famously {\em not} the unlink).

The above examples have led Low to formulate the following

\begin{Con} \label{Lowsconj}
Let $(M,g)$ be an orientable time oriented globally hyperbolic $(2+1)$-dimensional spacetime with a Cauchy surface $\Sigma \subset M$ diffeomorphic to a subset of $\bbR^2$; then $x,y \in M$ are causally related \emph{iff} $X,Y \subset N$ either intersect or are linked. 
\end{Con}

Some partial results have been achieved (see \cite{L88}, \cite{N00}, \cite{N02}), but the general case of this conjecture remains unproven.
%
%
%
\section{Skies in (2+1)-dimensional spacetimes} \label{section4}
As we saw in section \ref{section2}, giving a sky $X\subset N$ is equivalent to giving its cooriented wavefront $\pi(X) \subset \Sigma$. A natural question to ask is therefore what kinds of wavefronts can arise from projecting skies. More precisely, we make the following

\begin{Def}
A curve $\Phi \subset \bbR^2$ is said to be a {\em sky} if there exists an orientable time oriented globally hyperbolic $(2+1)$-dimensional spacetime $(M,g)$ with Cauchy surface $\Sigma \subset M$ diffeomorphic to a connected open subset $U \subset \bbR^2$ containing $\Phi$ such that $\Phi$ is (the image under the diffeomorphism of) the wavefront $\pi(X)$ generated at $\Sigma$ by some event $x \in M$.
\end{Def}

(Notice that we are using the same designation for skies and their projections on $\bbR^2$). The question above would then be more precisely phrased by asking which curves $\Phi \subset \bbR^2$ are skies. Notice that the projection $\pi:N \to \Sigma$ depends on the choice of $\Sigma$. Alternatively, one can think that $\pi$ is fixed and that changing $\Sigma$ corresponds to applying a diffeomorphism to $N$. We will be interested only in {\em generic} skies, i.e., those which are stable under small perturbations of $\Sigma$.

\begin{Def}
A {\em generic wavefront} is (the image of) a piecewise immersion of $S^1$ in $\bbR^2$ whose singularities are a finite even number of cusps, such that all points have multiplicity $1$ except for a finite number of points whith multiplicity $2$, where the self-intersection is transverse.
\end{Def}

Notice that if $\Phi \subset \bbR^2$ is a generic wavefront then it is always possible to define a coorientation on $\Phi$, i.e., a continuous unit normal vector field ${\bf n}:\Phi \to S^1$: This is true locally, since the only singularities are cusps; the fact that there exist an even number of cusps guarantees that this is also true globally. Equivalently, generic wavefronts can be lifted to Legendrian submanifolds of $N$ diffeomorphic to $S^1$. However, not all such manifolds project down to generic wavefronts, a meridian being a prime example; we shall call the projection of a Legendrian $S^1$ simply a {\em wavefront}. From this point on we shall often not distinguish between a cooriented wavefront and the corresponding Legendrian $S^1$.

\begin{Theo}
Generic skies are generic wavefronts.
\end{Theo}

\begin{proof}
This is a consequence of the fact that skies are projections of Legendrian submanifolds (see for instance \cite{EN00}).
\end{proof}

A further restriction to the class of curves which are allowed to be skies is topological in nature: it has to do with the fact that if $X \subset N$ is the sky of $x \in M$ then moving $\Sigma$ in $M$ (for instance using the flow of $\nabla^a t$) to a Cauchy surface through $x$ can be thought of as applying a Legendrian isotopy to $N$ which moves $X$ to a meridian. This shows that $X$ is in the same homology class of any meridian in $N$. In terms of the wavefront $\pi(X)$, this means that its coorientation must rotate by $2\pi$ as one traverses it in the positive direction.

\begin{Def}
The {\em winding number} of an oriented generic wavefront $\Phi$ is the integer $i(\Phi)$ such that the coorientation rotates by $2\pi i(\Phi)$ as $\Phi$ is traversed in the positive direction.
\end{Def}

Thus skies must have winding number $i=1$.

\begin{figure}[h]
\begin{center}    
        \includegraphics[scale=.4]{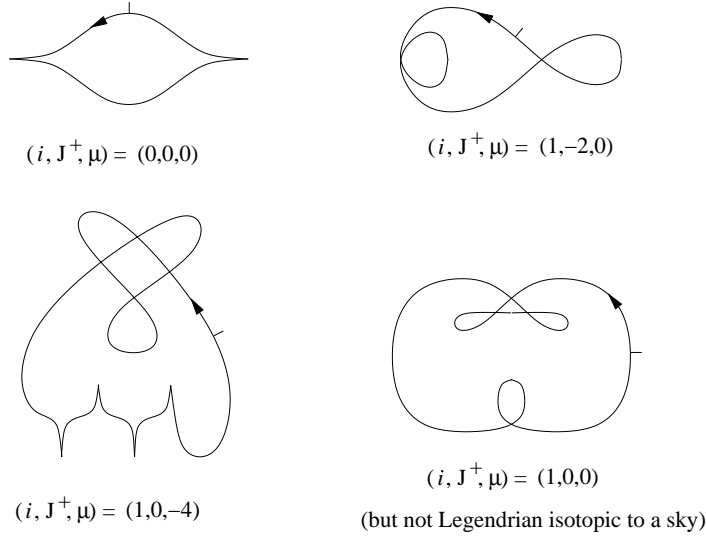}
\end{center}
\caption{Wavefronts not Legendre isotopic to skies.}
\label{picture2}
\end{figure}

The argument above actually shows that there exists a Legendrian isotopy carrying each sky to a meridian. This means that if a wavefront $\Phi\subset\bbR^2$ is a sky then all its Legendrian invariants must be the same as the Legendrian invariants of a meridian. Examples of such invariants are the {\em Maslov index} $\mu$ and the {\em Arnold} $J^+$ {\em invariant} (see \cite{Ar94}), whose values for a meridian are both $0$. In figure \ref{picture2} we show 3 examples of wavefronts which fail to have the correct values of the invariants $(i,J^+, \mu)$. It is a consequence of a theorem by Eliashberg (see \cite{EF98}) that in $S^3$ with the usual contact structure all unknotted Legendrian submanifolds diffeomorphic to $S^1$ having the same values of $(J^+,\mu)$ are Legendre isotopic; the same however is not true in our case, a counter-example being presented in the fourth wavefront in figure \ref{picture2}. This curve has the correct values of the invariants $(i,J^+, \mu)$, but we compute its Legendrian HOMFLY polynomial (see \cite{CG97}) to be $z_1\left(x^2+x^2y\right)+yz_1z_2$, as opposed to $z_1$ for a meridian.

\begin{figure}[h]
\begin{center}
        \includegraphics[scale=.4]{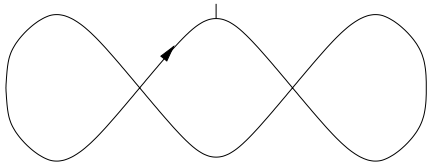}
\end{center}
\caption{Wavefront Legendre isotopic to a sky but not a sky.}
\label{picture3}
\end{figure}

It is not sufficient that $\Phi \subset \bbR^2$ is Legendre isotopic to a sky for it to be a sky. In fact, if $\Phi=\pi(X)$ then $\Phi$ must bound the plane region given by all points in $\Sigma$ which are causally related to $x\in M$. In particular, in all points of the curve which can be reached by a continuous curve coming from infinity (which we shall call the {\em outer boundary}) the coorientation must point in the same direction with respect to the curve. Thus for instance the plane curve in figure \ref{picture3} cannot be a sky. However this curve is Legendre isotopic to a circle (which is a sky): it is easy to see that any deformation of a wavefront such that the coorientation of different branches at self-intersection points never coincides (or equivalently, such that the coorientations of different branches have opposite directions at any self-tangency \footnote{Arnold calls these tangencies {\em safe self-tangencies}, by opposition to {\em dangerous self-tangencies}, which would correspond to self-intersections of the corresponding Legendrian $S^1$s; see \cite{Ar94}.}) induces a Legendrian motion of the corresponding Legendrian $S^1$.

\begin{figure}[h]
\begin{center}
        \includegraphics[scale=.4]{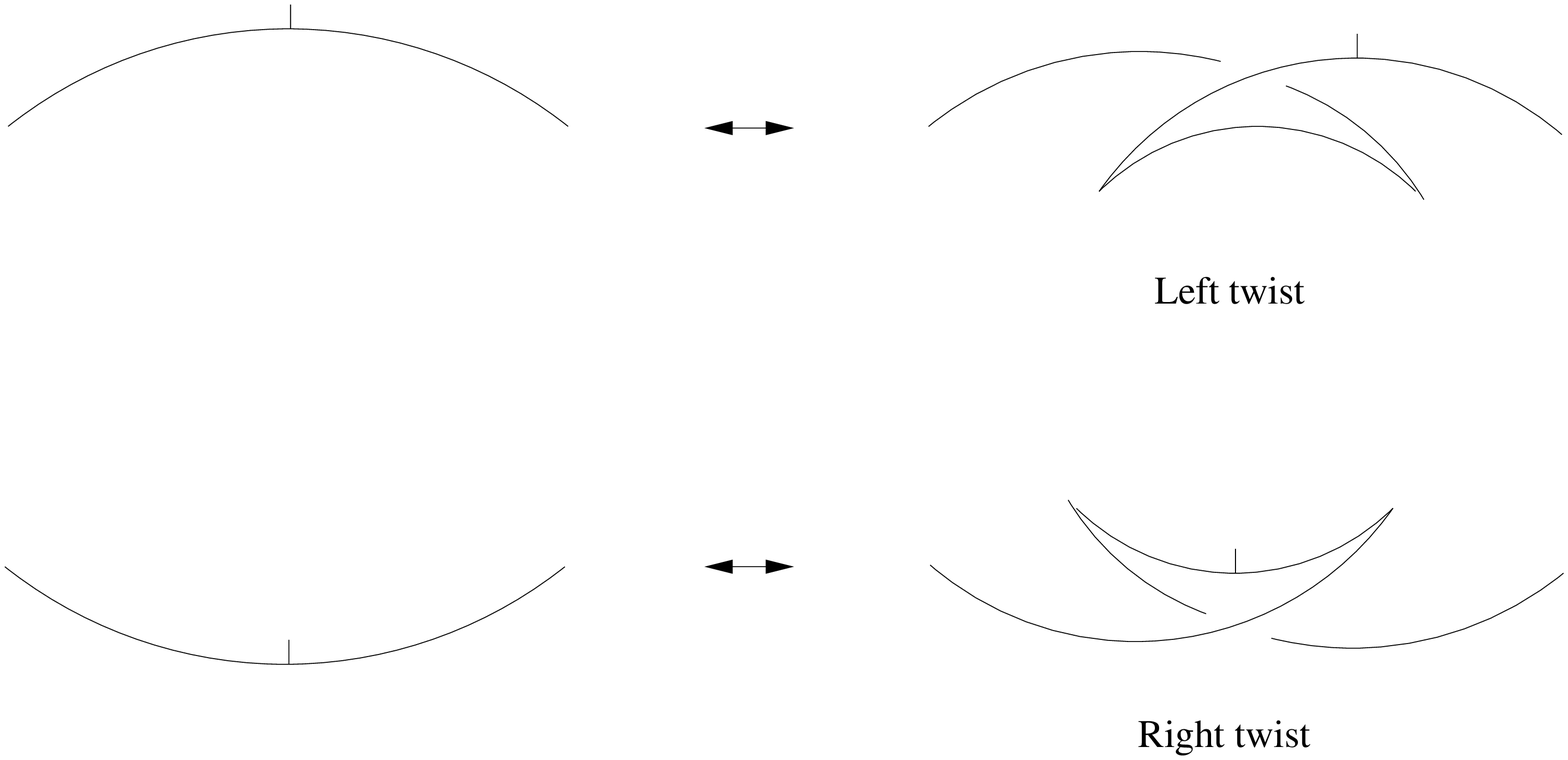}
\end{center}
\caption{Left and right twists.}
\label{picture4}
\end{figure}

The problem of deciding whether a given curve is a sky is therefore nontrivial. Instead of dwelling on this problem, let us assume that we are given a (generic) sky $\Phi=\pi(X)$, i.e., the wavefront generated at $\Sigma$ by $x \in M$. Consider a timelike curve $x_t$ such that $x_0 \in \Sigma$, $x_1 = x$, and let $\Phi_t=\pi(X_t)$ be the wavefront generated by $x_t$ at $\Sigma$. Then $\Phi_0$ is a point and $\Phi_t$ constitutes a Legendrian deformation of a point into $\Phi_1=\Phi$. The fact that it is a Legendrian deformation implies that the only changes allowed to $\Phi_t$ aside from moving (eventually crossing itself) are the creation or the destruction of a pair of cusps (see \cite{CG97} and figure \ref{picture4}). The pair of cusps is called a \emph{left twist} or a \emph{right twist} according to its position with respect to the coorientation, as shown in figure \ref{picture4}. The reason for these names is shown in the figure: these segments of wavefront are the projections of left or right twists of the corresponding Legendrian knots.

\begin{Theo}
As $\Phi_t$ evolves from $\Phi_0$ to $\Phi$ only left twists can be created and only right twists can be destroyed.
\end{Theo}

\begin{proof} Huygens's principle implies that right twists cannot be created on the portions of $\Phi_t$ which are part of the boundary of the causal set of $x_t$. Because the creation of a right twist is a local issue, we can always assume that the segment of wavefront where it is created is actually part of such a boundary. Thus the evolution cannot create right twists. The destruction of a left twist can be seen as the time-reversed creation of a right twist; if it were allowed, it would be possible to construct an evolution creating a right twist. Thus it cannot occur.
\end{proof}

The outer boundary of a generic sky carries a finite number of double points. Each double point is the projection of two points in the sky, which split the sky into two subsets, one of them projecting onto the outer boundary.

\begin{Def}
We shall call the subset of a sky arising from a double point in the outer boundary in this way and not projecting onto the outer boundary a \emph{pendant}.
\end{Def}

\begin{figure}[h]
\begin{center}
        \includegraphics[scale=.4]{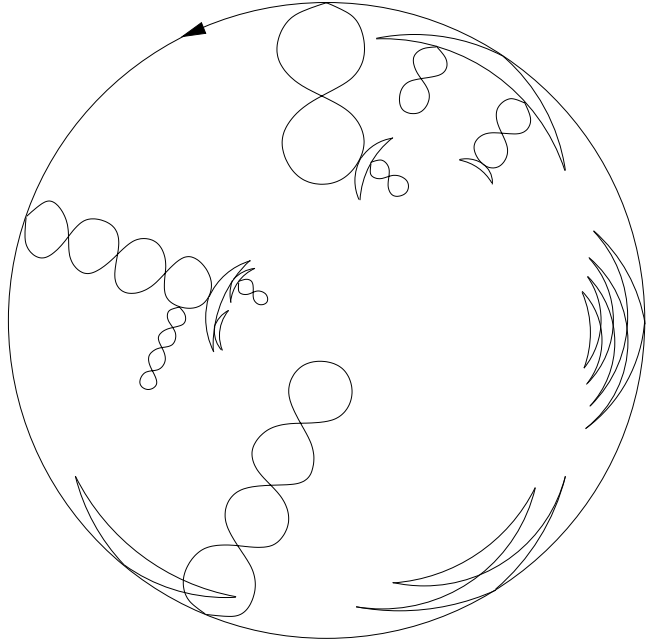}
\end{center}
\caption{Generic sky.}
\label{picture5}
\end{figure}

A generic sky can therefore be thought of as a circle (the outer boundary) with a finite number of pendants attached. The two basic mechanisms for forming pendants are overlaps (as for instance in Schwarzschild's (2+1)-dimensional spacetime) and growing cusps. Whenever a cusp forms, both sides of it are causally related to $x_t$; whenever an overlap forms a region not in the causal future may appear inside the outer boundary (as is the case in Schwarzschild's (2+1)-dimensional spacetime). Obviously one can use the same reasoning to divide a pendant into subpendants; in that respect a pendant is like a tree. See figure \ref{picture5} for an example.
%
%
%
\section{Linking and causality in (2+1)-dimensional spacetimes} \label{section5}
In this section we use the Kauffman polynomial to prove conjecture \ref{Lowsconj} for a certain class of skies of causally related points, of which infinitely many yield links with zero linking number.

Let $x,y \in M$ be two events not on the same null geodesic which are causally related. Assume that $y \in \Sigma$. Heuristically, $X$ and $Y$ must be linked because as $\Phi_t \subset \Sigma$ evolves from $\Phi_0=\{x_0\}$ to $\Phi_1 = \Phi$ it must hit $y$ at least once, and thus $y$ is inside the outer boundary of $\Phi$. If it is not in a pendant, then $\link(X,Y)=1$; if it is in a pendant where no left twisting has occurred, then $\link(X,Y)\geq 1$; if it is in a pendant where left twisting has occurred, and hence possibly $\link(X,Y)=0$, $X \sqcup Y$ cannot be deformed to the unlink because there is no right twisting anywhere in $X$ to ``cancel off'' the left twisting.

Unfortunately, it is not easy to turn this insight into a proof of Low's conjecture. However it can be used to prove Low's conjecture for a large class of examples.

\begin{figure}[h]
\begin{center}
        \includegraphics[scale=.4]{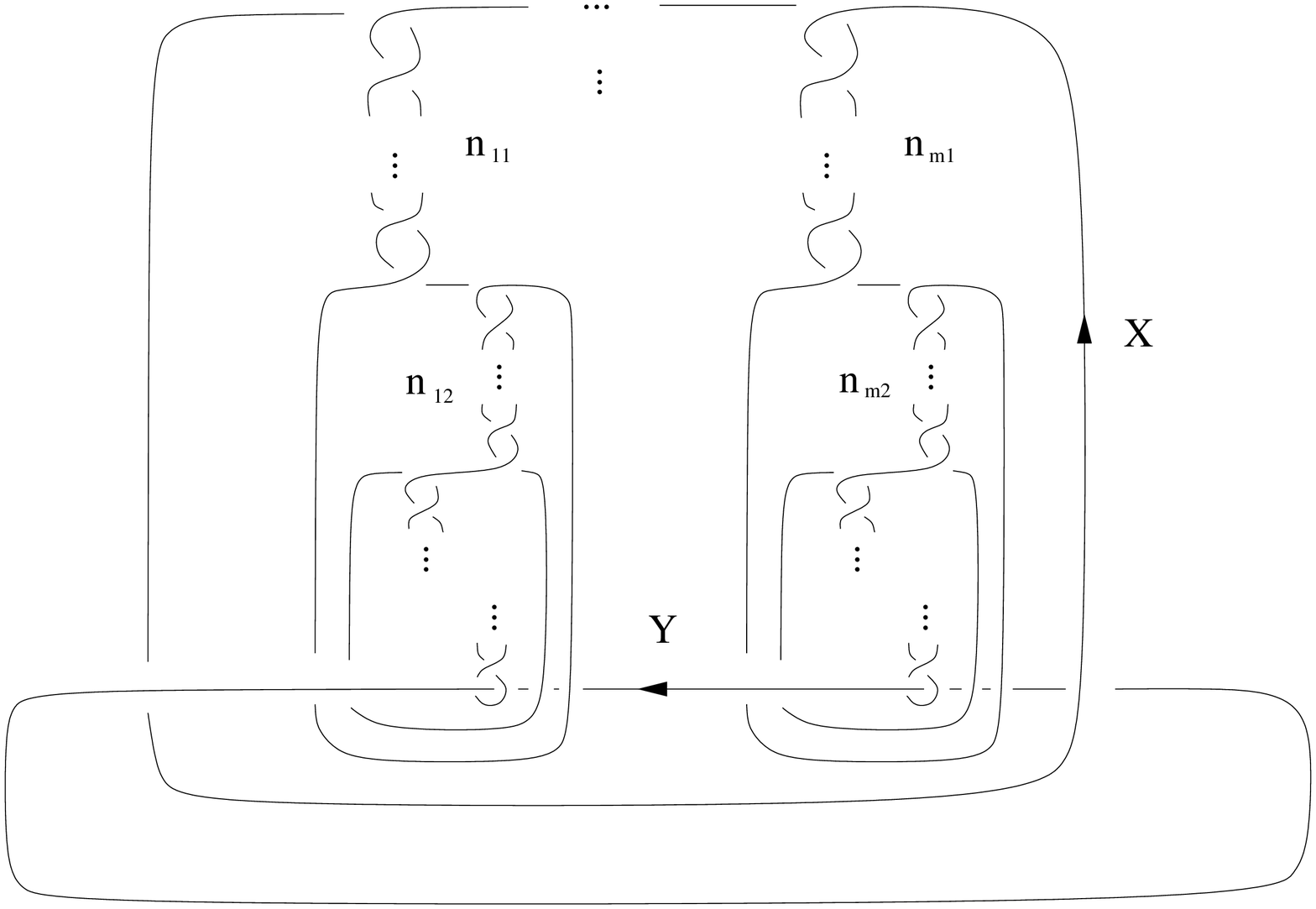}
\end{center}
\caption{Sky of a given type.}
\label{picture6}
\end{figure}

\begin{Def}
Let $n_{11},...,n_{1k_1},n_{21},...,n_{2k_2},...,n_{m1},...,n_{mk_m}\in \bbN$. Let $x,y\in M$. Then $(X,Y)$ is said to be a \emph{pair of skies of type} 
\[
\left( 
\begin{array}{cccc}
n_{11} & n_{21} & ... & n_{m1} \\ 
... & ... & ... & ... \\ 
n_{1k_1} & n_{2k_2} & ... & n_{mk_m}
\end{array}
\right) 
\]
if $X \sqcup Y$ admits a link diagram as shown if figure \ref{picture6} (when embedded in $\bbR^3$ through the standard diffeomorphism; each integer refers to the number of double points).
\end{Def}

It is easy to see how such skies may form, each double point of $X$ corresponding to two cusps in $\Phi$. Thus for instance the pair of skies with linking number 0 we considered in section \ref{section3} is a pair of skies of type $\left( 2 \right)$.

We now prove that such skies are always linked. In order to do this we must recall the following

\begin{figure}[h]
\begin{center}
        \includegraphics[scale=.4]{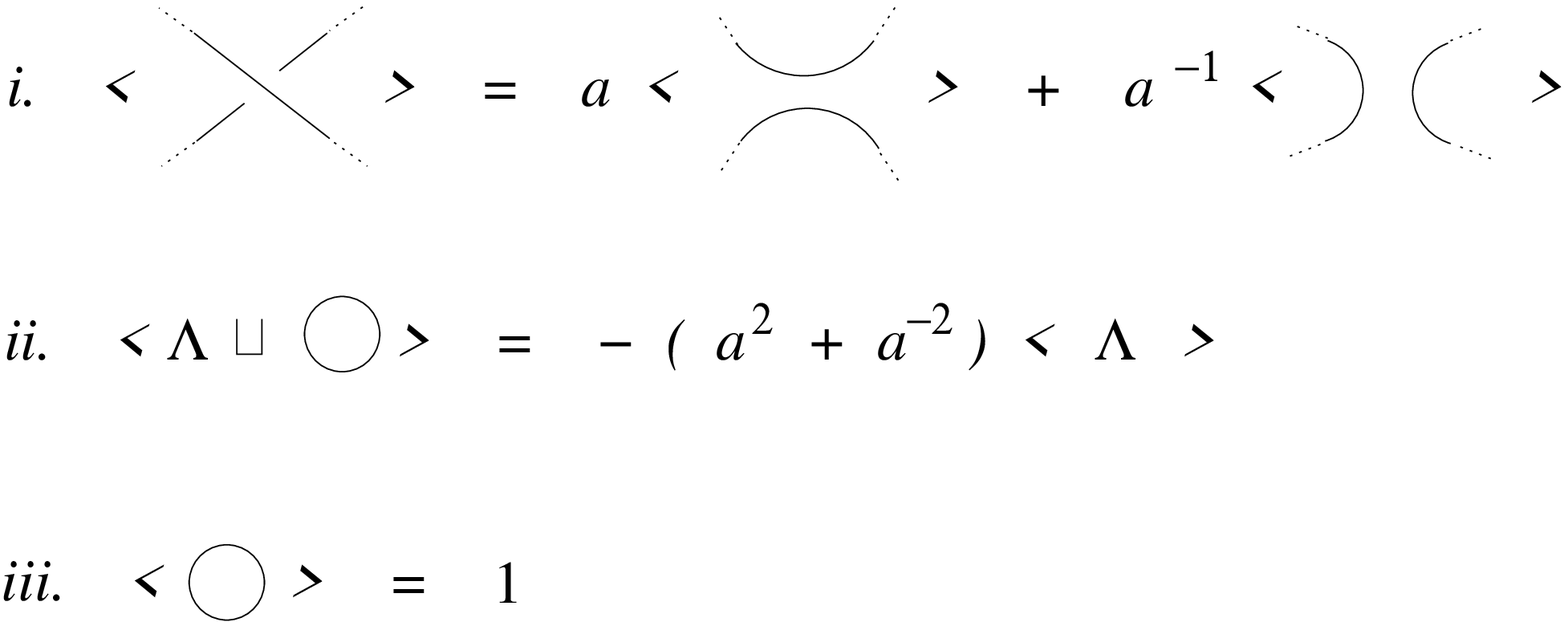}
\end{center}
\caption{Rules defining the bracket polynomial.}
\label{picture7}
\end{figure}

\begin{Def}
The {\em bracket polynomial} $\langle \Lambda \rangle$ of a link diagram $\Lambda$ is the polynomial in the variables $a^{\pm 1}$ defined by the rules shown in figure \ref{picture7} (see \cite{PS91}).
\end{Def}

Notice that the bracket polynomial does not depend on the orientations of the link, a fact that we will use to our advantage. The bracket polynomial is {\em not} a link invariant, but can be used to build a link invariant. In order to do so we must introduce the following

\begin{figure}[h]
\begin{center}
        \includegraphics[scale=.4]{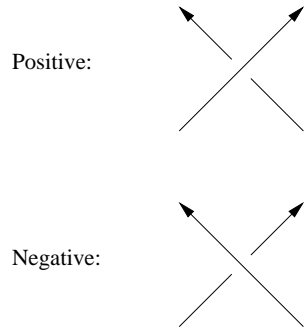}
\end{center}
\caption{Positive and negative crossings.}
\label{picture8}
\end{figure}

\begin{Def}
A crossing in a link diagram is said to be {\em positive} or {\em negative} according to whether the branch going rightward goes over or under the branch going leftward (right and left being defined with respect to the orientations of the branches; see figure \ref{picture8}).
\end{Def}

\begin{Def}
The {\em writhe number} $w(\Lambda )$ of a link diagram $\Lambda $ is the sum of the signs of all crossings (where the sign of a crossing is $\pm 1$ according to whether it is positive or negative).
\end{Def}

\begin{Def}
The \emph{Kauffman polynomial} of a link $L$ is 
\[
K\left( L\right) =\left( -a\right) ^{-3w(\Lambda )}\langle\Lambda \rangle
\]
where $\Lambda $ is any link diagram for $L$.
\end{Def}

\begin{Theo}
The Kauffman polynomial is a link invariant.
\end{Theo}

\begin{proof}
See \cite{PS91}.
\end{proof}

The Kauffman polynomial is related to the Jones polynomial in one variable through the variable change $a=q^{-\frac 14}$.

To prove that skies of the type we are considering are always linked we compute certain invariants obtained from the Kauffman polynomial. In order to do so we shall need the following

\begin{figure}[h]
\begin{center}
        \includegraphics[scale=.4]{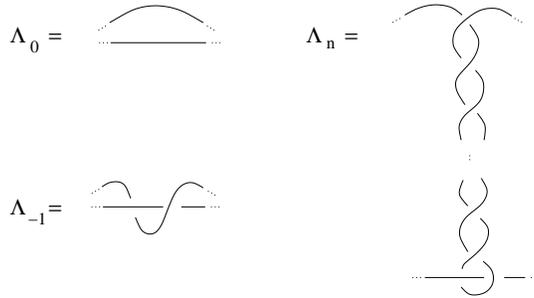}
\end{center}
\caption{Basic link diagrams.}
\label{picture9}
\end{figure}

\begin{Theo} \label{formula}
Let $\Lambda _n$, $\Lambda _0$, $\Lambda _{-1}$ be as in figure \ref{picture9} ($n$ being the number of double points in the twisted knot). Then 
\[
\left\langle \Lambda _n\right\rangle =a^n\left\langle \Lambda_{-1}\right\rangle -\left( a^4+a^{-4}\right) \sum_{i=1}^n(-1)^{-3(n-i)}a^{-3n+4i-2}\left\langle \Lambda _0\right\rangle.
\]
\end{Theo}

\begin{figure}[h]
\begin{center}
        \includegraphics[scale=.4]{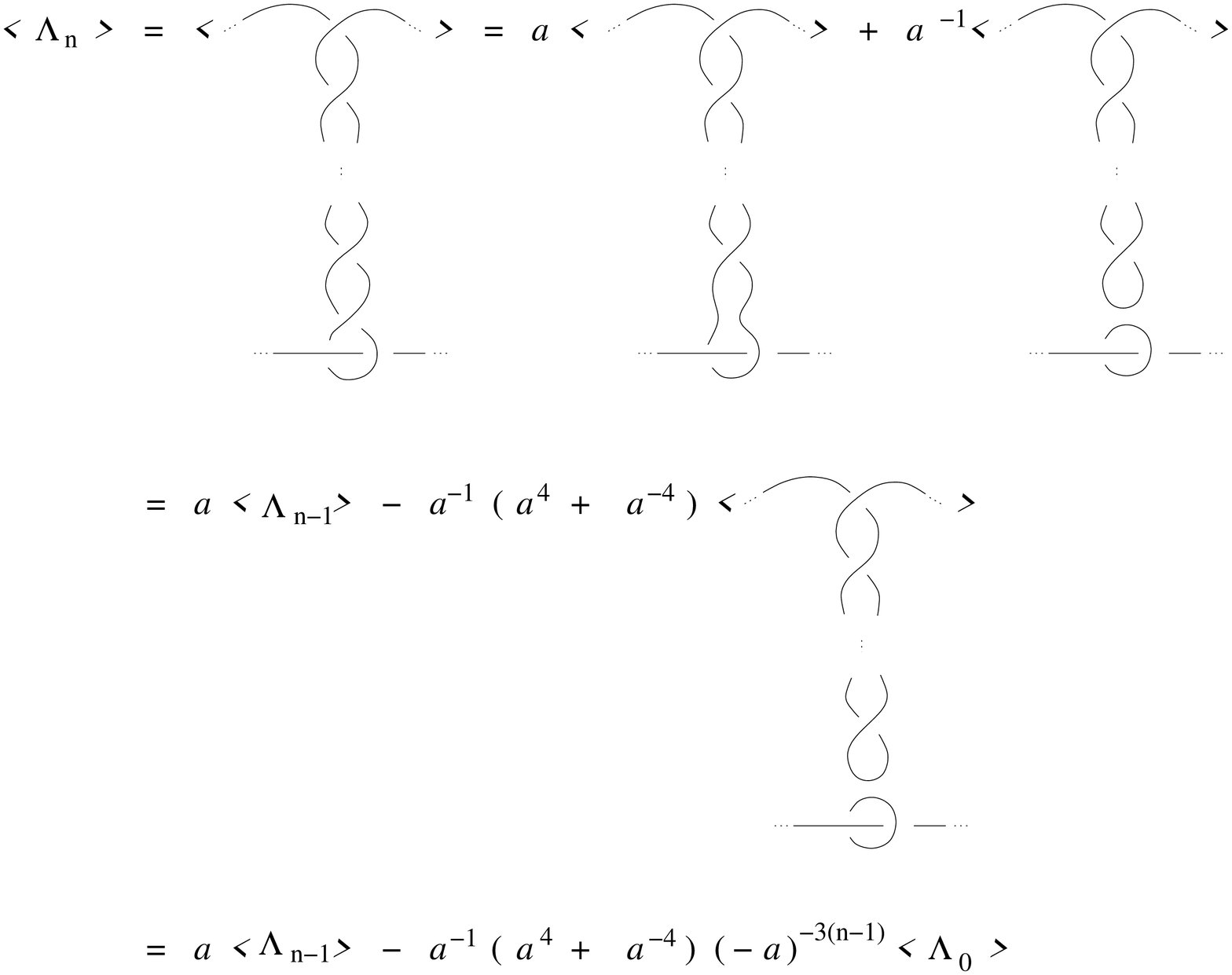}
\end{center}
\caption{Proof of the inductive step of theorem \ref{formula}}
\label{picture11}
\end{figure}

\begin{proof}
We prove this result by induction. For $n=1$ all there is to prove is 
\[
\left\langle \Lambda _1\right\rangle =a\left\langle \Lambda_{-1}\right\rangle -\left( a^4+a^{-4}\right) a^{-1}\left\langle \Lambda_0\right\rangle .
\]

This is a simple application of the rules defining the bracket polynomial and is done in figure \ref{picture10}.

\begin{figure}[h]
\begin{center}
        \includegraphics[scale=.4]{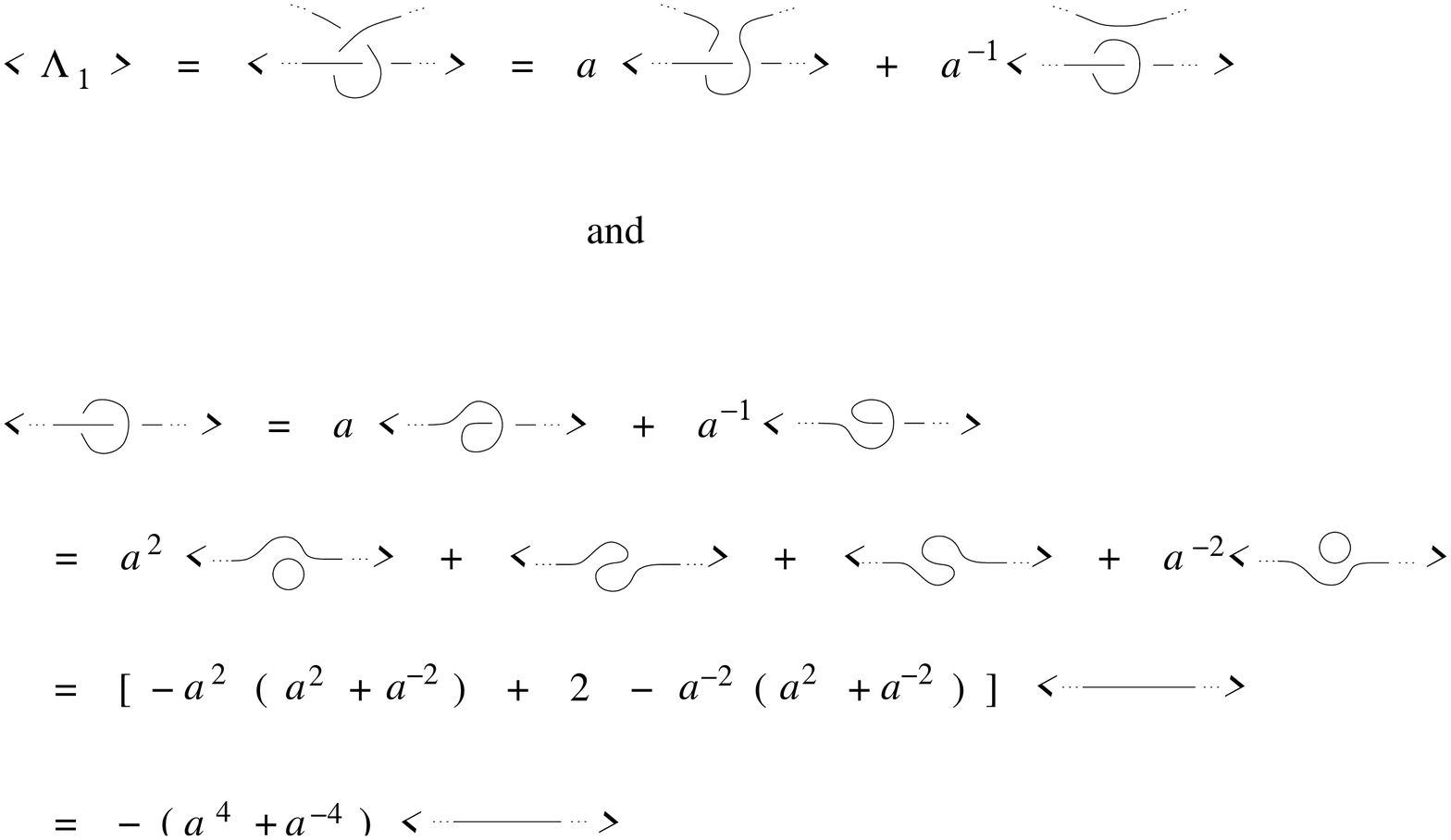}
\end{center}
\caption{Proof of theorem \ref{formula} for $n=1$.}
\label{picture10}
\end{figure}

As for the inductive step, our formula yields

\begin{eqnarray*}
\left\langle \Lambda _n\right\rangle &=&aa^{n-1}\left\langle \Lambda_{-1}\right\rangle -\left( a^4+a^{-4}\right)\sum_{i=2}^n(-1)^{-3(n-i)}a^{-3n+4i-2}\left\langle \Lambda _0\right\rangle \\
&&-\left( a^4+a^{-4}\right) (-1)^{-3(n-1)}a^{-3n+2}\left\langle \Lambda_0\right\rangle \\
&=&a\left[ a^{n-1}\left\langle \Lambda _{-1}\right\rangle -\left(a^4+a^{-4}\right)\sum_{j=1}^{n-1}(-1)^{-3(n-1-j)}a^{-3(n-1)+4j-2}\left\langle \Lambda_0\right\rangle \right] \\
&&-a^{-1}\left( a^4+a^{-4}\right) (-a)^{-3(n-1)}\left\langle \Lambda_0\right\rangle \\
&=&a\left\langle \Lambda _{n-1}\right\rangle -a^{-1}\left( a^4+a^{-4}\right)(-a)^{-3(n-1)}\left\langle \Lambda _0\right\rangle
\end{eqnarray*}

and hence all that must be proved is 

\[
\left\langle \Lambda _n\right\rangle =a\left\langle \Lambda_{n-1}\right\rangle -a^{-1}\left( a^4+a^{-4}\right)(-a)^{-3(n-1)}\left\langle \Lambda _0\right\rangle.
\]

This can be done much as the $n=1$ case; we do so in figure \ref{picture11}. The last step in the proof is justified in lemma \ref{writhe}.
\end{proof}

\begin{figure}[h]
\begin{center}
        \includegraphics[scale=.4]{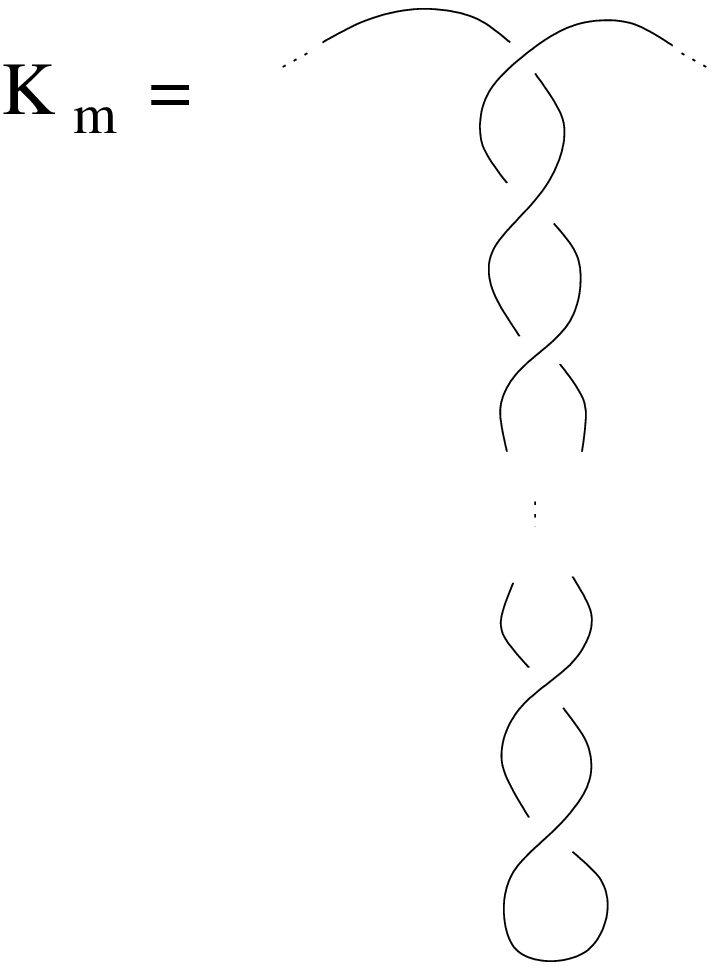}
\end{center}
\caption{Definition of $K_m$.}
\label{picture12}
\end{figure}

\begin{Lemma}\label{writhe}
Let $K_m$ be the link diagram shown in figure \ref{picture12} ($m$ being the number of double points). Then $\left\langle K_m\right\rangle =\left( -a\right)^{-3m}\left\langle K_0\right\rangle$.
\end{Lemma}

\begin{proof}
Choose any orientation on $K_m$ (and hence on $K_0$). Clearly $w\left( K_m\right) =w\left( K_0\right) -m$. Since $K_m$ and $K_0$ are link diagrams for the same link, they must have the same Kauffman polynomial, and hence 
\begin{eqnarray*}
K\left( K_m\right) = K\left( K_0\right) &\Leftrightarrow & \left( -a\right)^{-3w\left( K_m\right) }\left\langle K_m\right\rangle =\left( -a\right)^{-3w\left( K_0\right) }\left\langle K_0\right\rangle \\
&\Leftrightarrow &\left( -a\right) ^{-3w\left( K_0\right) +3m}\left\langle K_m\right\rangle =\left( -a\right) ^{-3w\left( K_0\right) }\left\langle K_0\right\rangle \\
&\Leftrightarrow &\left\langle K_m\right\rangle =\left( -a\right)^{-3m}\left\langle K_0\right\rangle.
\end{eqnarray*}

\end{proof}

Notice that it is in this lemma that the left twisting comes into the proof.

We will be only interested in the terms of higher and lower order of bracket polynomials. It is therefore useful to keep in mind the slightly simpler (if less accurate) formula given in theorem \ref{formula}: 

\begin{equation} \label{fundamental}
\left\langle \Lambda _n\right\rangle =a^n\left\langle \Lambda_{-1}\right\rangle \pm a^{n+2}\left\langle \Lambda _0\right\rangle \pm ...\pm a^{-3n-2}\left\langle \Lambda _0\right\rangle 
\end{equation}

\begin{Def}
Let $n_{11},...,n_{1k_1},n_{21},...,n_{2k_2},...,n_{m1},...,n_{mk_m}\in \Bbb{%
N}$. We shall denote by 
\[
\left( 
\begin{array}{cccc}
n_{11} & n_{21} & ... & n_{m1} \\ 
... & ... & ... & ... \\ 
n_{1k_1} & n_{2k_2} & ... & n_{mk_m}
\end{array}
\right) 
\]
the standard link diagram for the link of such type (depicted in figure \ref{picture6}). Also, we shall also allow $n_{ik_i}$ to assume the values 0,-1 with the meaning described in figure \ref{picture9}.
\end{Def}

\begin{Theo}
Let $n_{11},...,n_{1k_1},n_{21},...,n_{2k_2},...,n_{m1},...,n_{mk_m}\in \bbN$. Then 
\[
\left\langle \left( 
\begin{array}{cccc}
n_{11} & n_{21} & ... & n_{m1} \\ 
... & ... & ... & ... \\ 
n_{1k_1} & n_{2k_2} & ... & n_{mk_m}
\end{array}
\right) \right\rangle = \pm a^{N+2k+4}\pm ...\pm a^{-3N-2k-4}
\]
where 
\[
N=\sum_{i=1}^m\sum_{j=1}^{k_i}n_{ij}
\]
is the total number of double points and 
\[
k=\sum_{i=1}^mk_i
\]
is the total number of subpendants of the knot diagram of $X$.
\end{Theo}

\begin{proof}
It should be clear that $\left( 0 \right) $ is a link diagram for the Hopf link and $\left( -1\right) $ is a link diagram for the unlink, and hence 
\begin{eqnarray*}
\left\langle \left( 0\right) \right\rangle &=&-\left( a^4+a^{-4}\right); \\
\left\langle \left( -1\right) \right\rangle &=&-\left( a^2+a^{-2}\right).
\end{eqnarray*}

Consequently formula \ref{fundamental} yields 
\begin{eqnarray*}
\left\langle \left( n\right) \right\rangle &=&a^n\left\langle \left(-1\right) \right\rangle \pm a^{n+2}\left\langle \left( 0\right) \right\rangle \pm ...\pm a^{-3n-2}\left\langle \left( 0\right) \right\rangle\\
&=&a^{n+6}\pm ...\pm a^{-3n-6}.
\end{eqnarray*}

and the theorem's conclusion holds in this case.

Next, we notice that 
\[
\left( 
\begin{array}{c}
n_1 \\ 
0
\end{array}
\right) = \left( n_1\right) \Rightarrow \left\langle \left( 
\begin{array}{c}
n_1 \\ 
0
\end{array}
\right) \right\rangle = a^{n_1+6}\pm ...\pm a^{-3n_1-6} 
\]
and that lemma \ref{writhe} yields 
\begin{eqnarray*}
\left\langle \left( 
\begin{array}{c}
n_1 \\ 
-1
\end{array}
\right) \right\rangle &=&\left( -a\right) ^{-3n_1}\left\langle \left(0\right) \right\rangle \\
&=&-\left( -a\right) ^{-3n_1}\left( a^4+a^{-4}\right) \\
&=&\pm a^{-3n_1+4}\pm a^{-3n_1-4}.
\end{eqnarray*}

Consequently, one can ignore the corresponding term in formula \ref{fundamental}, 
\[
\left\langle \left( 
\begin{array}{c}
n_1 \\ 
n_2
\end{array}
\right) \right\rangle = \pm a^{n_1+n_2+8}\pm ...\pm a^{-3n_1-3n_2-8}.
\]
and again the theorem's conclusion holds. In general, one has 
\[
\left( 
\begin{array}{c}
n_1 \\ 
... \\ 
n_{k-1} \\ 
0
\end{array}
\right) = \left( 
\begin{array}{c}
n_1 \\ 
... \\ 
n_{k-1}
\end{array}
\right) 
\]
and consequently 
\[
\left\langle \left( 
\begin{array}{c}
n_1 \\ 
... \\ 
n_{k-1} \\ 
0
\end{array}
\right) \right\rangle = \pm a^{n_1+...+n_{k-1}+2k+2}\pm ...\pm a^{-3n_1-...-3n_{k-1}-2k-2}.
\]

From lemma \ref{writhe}, on the other hand, one gets 
\begin{eqnarray*}
\left\langle \left( 
\begin{array}{c}
n_1 \\ 
... \\ 
n_{k-1} \\ 
-1
\end{array}
\right) \right\rangle &=& \left( -a\right) ^{-3n_{k-1}}\left\langle \left( 
\begin{array}{c}
n_1 \\ 
... \\ 
n_{k-2}
\end{array}
\right) \right\rangle \\
&=&\pm a^{n_1+...+n_{k-2}-3n_{k-1}+2k}\pm ...\pm a^{-3n_1-...-3n_{k-1}-2k}
\end{eqnarray*}
and again one can ignore the corresponding term in formula \ref{fundamental}, thus getting 
\[
\left\langle \left( 
\begin{array}{c}
n_1 \\ 
... \\ 
n_{k-1}
\end{array}
\right) \right\rangle =\pm a^{n_1+...+n_k+2k+4}\pm ...\pm a^{-3n_1-...-3n_k-2k-4}.
\]

This proves the theorem for $m=1$. For $m>1$ the proof easily follows (by using the exact same reasoning) from the observation that 
\[
\left( 
\begin{array}{cccc}
n_{11} & ... & n_{m1} & 0 \\ 
... & ... & ... &  \\ 
n_{1k_1} & ... & n_{mk_m} & 
\end{array}
\right) =\left( 
\begin{array}{cccc}
n_{11} & ... & n_{m1} & -1 \\ 
... & ... & ... &  \\ 
n_{1k_1} & ... & n_{mk_m} & 
\end{array}
\right) =\left( 
\begin{array}{ccc}
n_{11} & ... & n_{m1} \\ 
... & ... & ... \\ 
n_{1k_1} & ... & n_{mk_m}
\end{array}
\right) 
\]
and that consequently one can go on ignoring the first term in formula \ref{fundamental}.
\end{proof}

\begin{Theo}
Let $n_{11},...,n_{1k_1},n_{21},...,n_{2k_2},...,n_{m1},...,n_{mk_m}\in \bbN$, and $(X,Y)$ be of type 
\[
\left( 
\begin{array}{cccc}
n_{11} & n_{21} & ... & n_{m1} \\ 
... & ... & ... & ... \\ 
n_{1k_1} & n_{2k_2} & ... & n_{mk_m}
\end{array}
\right).
\]
Then 
\[
K(X \sqcup Y) = \pm a^{4N+2k-6l+4}\pm ...\pm a^{-2k-6l-4}
\]
where 
\[
l=\link(X,Y).
\]
\end{Theo}

\begin{proof}
We just have to compute the writhe number of the standard link diagram for the link $X \sqcup Y$ (depicted in figure \ref{picture6}). There are two kinds of crossings in this link diagram: those involving only the knot diagram of $X$ ($N$ of them, all corresponding to negative crossings) and those involving the two knot diagrams. It is a well known fact that the sum of the signs of the latter crossings is equal to twice the linking number of $X$ and $Y$ (see \cite{R90}). Consequently, the writhe number is 
\[
w=-N+2l 
\]
and hence 
\begin{eqnarray*}
K(X\sqcup Y) &=&\left( -a\right) ^{-3w}\left\langle \left( 
\begin{array}{cccc}
n_{11} & n_{21} & ... & n_{m1} \\ 
... & ... & ... & ... \\ 
n_{1k_1} & n_{2k_2} & ... & n_{mk_m}
\end{array}
\right) \right\rangle \\
&=&\left( -a\right) ^{3N-6l}\left( \pm a^{N+2k+4}\pm ...\pm
a^{-3N-2k-4}\right) \\
&=&\pm a^{4N+2k-6l+4}\pm ...\pm a^{-2k-6l-4}.
\end{eqnarray*}
\end{proof}

\begin{Cor}
Let $n_{11},...,n_{1k_1},n_{21},...,n_{2k_2},...,n_{m1},...,n_{mk_m}\in \bbN$, and $(X,Y)$ be of type 
\[
\left( 
\begin{array}{cccc}
n_{11} & n_{21} & ... & n_{m1} \\ 
... & ... & ... & ... \\ 
n_{1k_1} & n_{2k_2} & ... & n_{mk_m}
\end{array}
\right).
\]
Then $X$ and $Y$ are linked.
\end{Cor}

\begin{proof}
If $l\neq 0$ then obviously $X$ and $Y$ are linked. If $l=0$ then 
\[
K(X \sqcup Y)=\pm a^{4N+2k+4}\pm ...\pm a^{-2k-4}\neq -a^2-a^{-2} 
\]
and consequently $X\sqcup Y$ is not the unlink.
\end{proof}

Since both $\link(X,Y)$ and the exponents of the terms of higher and lower order of $K\left( X\sqcup Y\right) $ are clearly invariants of $X\sqcup Y$, one can in fact conclude that

\begin{Cor}
Let $n_{11},...,n_{1k_1},n_{21},...,n_{2k_2},...,n_{m1},...,n_{mk_m}\in \bbN$, and $(X,Y)$ be of type 
\[
\left( 
\begin{array}{cccc}
n_{11} & n_{21} & ... & n_{m1} \\ 
... & ... & ... & ... \\ 
n_{1k_1} & n_{2k_2} & ... & n_{mk_m}
\end{array}
\right).
\]
Then 
\[
N=\sum_{i=1}^m\sum_{j=1}^{k_i}n_{ij}
\]
and 
\[
k=\sum_{i=1}^mk_i
\]
are invariants of $X\sqcup Y$.
\end{Cor}

\begin{figure}[h]
\begin{center}
        \includegraphics[scale=.4]{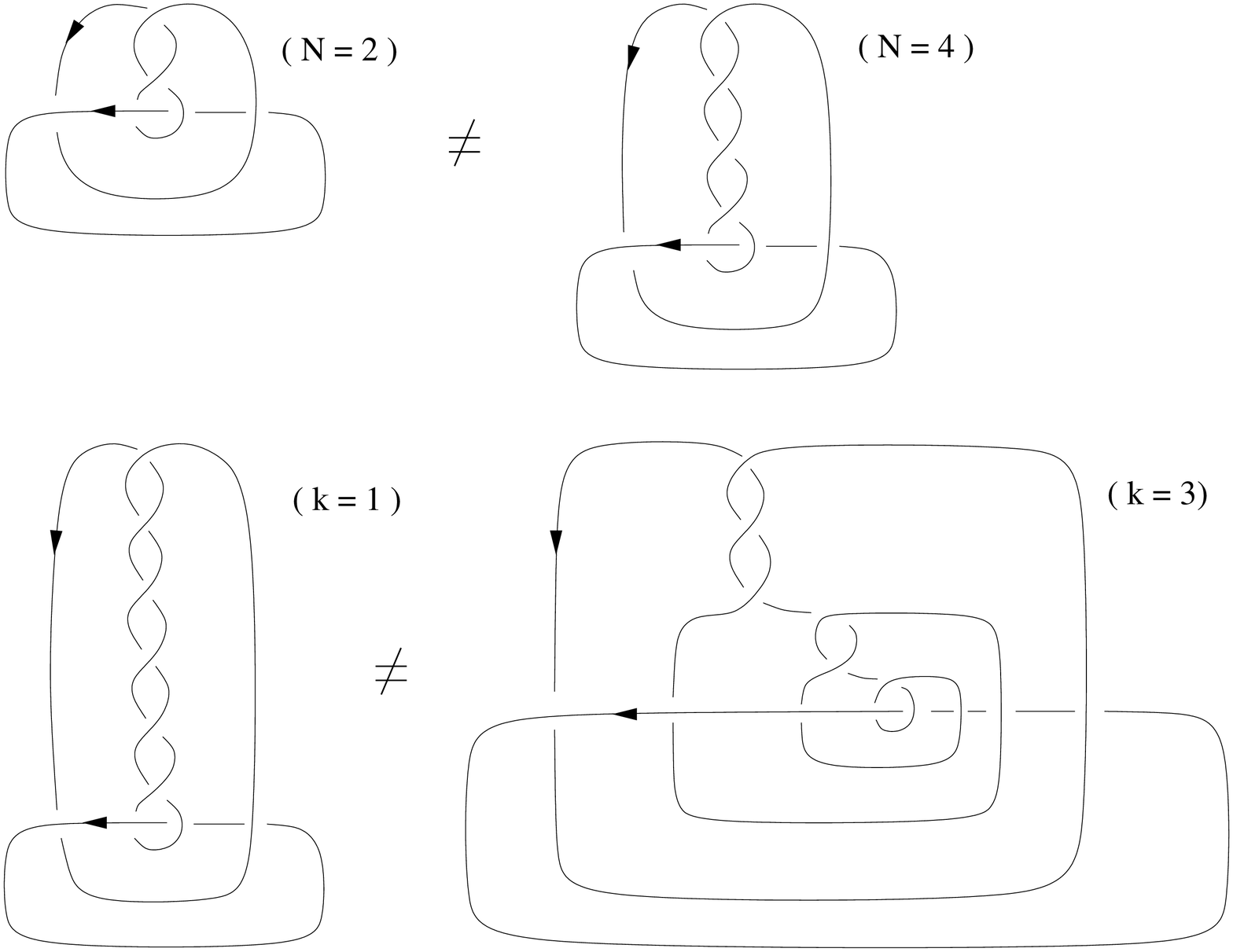}
\end{center}
\caption{Pairs of skies with zero linking number forming non-equivalent nontrivial links.}
\label{picture13}
\end{figure}

Thus not only are all skies of the type we have considered linked, but also pairs of skies with different values of $N$ and $k$ form non-equivalent links. See figure \ref{picture13} for some examples with zero linking number.

This method can be successfully employed in some generalizations of the skies we have considered. For instance, it can handle multiple subpendants and a pendant overlapping itself. Unfortunately, it is hard to see how it can handle \emph{generic} skies.

As we said, the bracket polynomial (and hence the Kauffman polynomial when calculated through it) is especially amenable to the kind of computations we have to do, because it is local and orientation-independent. Using the usual definitions of the Kauffman (or other) polynomial in terms of skein relations turns out to be less fruitful, as one must continuously worry about related knots and orientations. See \cite{L88} for such computations using the Conway polynomial.
%
%
%
\section{Linking and causality in (3+1)-dimensional spacetimes} \label{section6}
\begin{figure}[h]
\begin{center}
        \includegraphics[scale=.4]{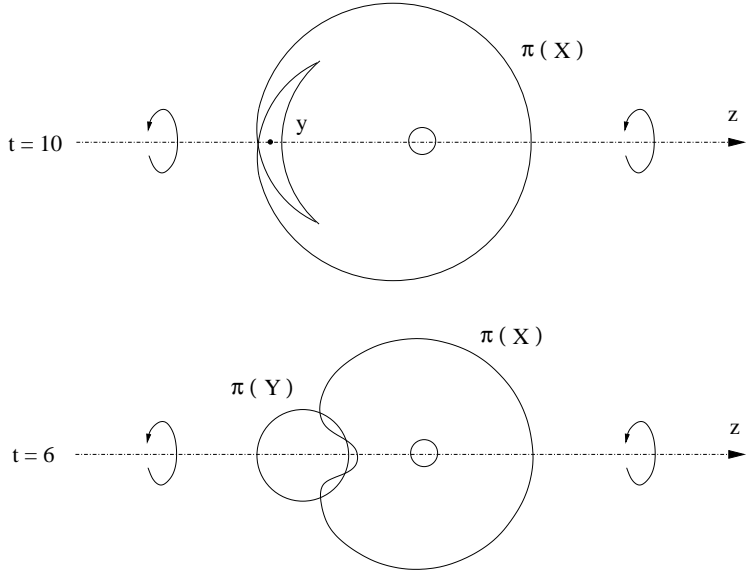}
\end{center}
\caption{Causally related events with unlinked skies.}
\label{picture14}
\end{figure}

As was noted by Low (\cite{L88}), in $(3+1)$-dimensional spacetimes it is possible to obtain causally related events whose skies are unlinked. A simple example is obtained by considering $\bbR^4$ endowed with the line element
\[
ds^2=dt^2-\Omega^2\left(x^1,x^2,x^3\right)\left(\left(dx^1\right)^2+\left(dx^2\right)^2+\left(dx^3\right)^2\right),
\]
where $\Omega:\bbR^3 \to [1, + \infty)$ is a smooth function equal to $1$ except on the ball $\left(x^1\right)^2+\left(x^2\right)^2 + \left(x^3\right)^2 < 1$, where it increases radially towards the center. The wavefront of the event $x = (0,0,0,4)$ on the Cauchy surface $t=10$ is the revolution surface generated by the curve depicted in figure \ref{picture14}. Take an event $y$ in the Cauchy surface as shown in the figure.

\begin{Prop} \label{unlink}
$X \sqcup Y$ is the unlink.
\end{Prop}

\begin{proof}
Move the Cauchy surface towards the past; the projections $\pi(X), \pi(Y)$ eventually change into the surfaces of revolution generated by the curves depicted in the lower diagram in figure \ref{picture14}). Recall that the intersection of, say, $X$, with a $S^2$ fiber is determined by the normal vector to $\pi (X)$ at that point. Because the Cauchy surface we are considering is to the future of $x$ but to
the past of $y$, the relevant normal is the outward-pointing normal for $\pi(X)$ and the inward-pointing normal for $\pi (Y)$. Let $(\rho ,\varphi ,z)$ be cylindrical coordinates in $\Bbb{R}^3$, and consider the deformation of $X$ given by 
\[
f_t \left(\xi ,{\bf n} \right) = \left(\xi ,\frac{{\bf n} + t \frac{\partial}{\partial \varphi}}{\left\| {\bf n} + t \frac{\partial}{\partial \varphi} \right\|}\right) 
\]
for $\xi \in \pi (X)$ and ${\bf n}$ the normal vector to $\pi (X)$ at $\xi$. It should be clear that one can move $X$ away from $Y$ (by moving $\pi \left(f_1 \left(X \right) \right)$ away from $\pi(Y)$ along the $z$-axis, say) without $X$ ever intersecting $Y$, and that once one has moved $X$ far enough both can be deformed to skies of points on the Cauchy surface (which by definition are unlinked).
\end{proof}

Notice that in the proof of proposition \ref{unlink} $f_t(X)$ is not a Legendrian knot for all $t>0$, and hence we cannot conclude that $X \sqcup Y$ is the Legendrian unlink.

Legendrian linking provides us with a more restrictive concept of linking which fits nicely with the natural contact structure of the manifold of light rays. An illustration of this restrictiveness is given by the following theorem by Lisa Traynor:

\begin{Theo} \label{Lisa}
Let $(M,g)$ be Minkowski (2+1)-spacetime and $N$ the corresponding (contact) manifold of light rays. Let $x,y\in M$ be causally related points not in the same null geodesic. Then $X\sqcup Y$ and $Y\sqcup X$ are \emph{not} equivalent Legendrian links.
\end{Theo}

\begin{proof}
See \cite{T97}.
\end{proof}

\begin{Cor}
Let $(M,g)$ be Minkowski (2+1)-spacetime and $N$ the corresponding (contact) manifold of light rays. Let $x,y,z \in M$ with $x$ causally related to $y$ and not in the same null geodesic, and not causally related to $z$. Then $X\sqcup Y$, $Y\sqcup X$ and $X\sqcup Z$ yield the different equivalence classes of Legendrian links formed by pairs of skies.
\end{Cor}

\begin{proof}
This is an immediate consequence of theorem \ref{Lisa} and the fact that any two points on $M$ not lying on the same null geodesic can be moved into $x$ and $y$, $y$ and $x$ or $x$ and $z$ in such a way that they never are on the same null geodesic.
\end{proof}

Note that if $x,y$ are causally related and not on the same null geodesic and (say) $t(x) < t(y)$ then $X\sqcup Y$ and $Y\sqcup X$ are clearly equivalent smooth links: one can always find coordinates on $M$ such that on  $N=\Bbb{R}^2\times S^1$ we have 
\begin{eqnarray*}
X &=&\left\{ (R\cos \theta ,R\sin \theta , \theta ): 0 \leq \theta \leq 2\pi \right \} \\
Y &=&\left\{ (R\cos (\theta + \pi ),R\sin (\theta + \pi), \theta ): 0 \leq \theta \leq 2\pi \right \}
\end{eqnarray*}
for some $R>0$; thus we see that 
\[
\Phi _t \left(x^1,x^2,\theta \right) = \left( \cos \left( \pi t\right)x^1-\sin \left( \pi t\right) x^2,\sin \left( \pi t\right) x^1+\cos \left(\pi t\right) x^2,\theta \right) 
\]
is a smooth isotopy which carries $X\sqcup Y$ to $Y\sqcup X$. Thus there are only two distinct equivalence classes of links formed by pairs of skies in Minkowski (2+1)-spacetime. This is related to the fact that in the (2+1)-dimensional case $\link(X,Y)=\link(Y,X)$.

It is instructive to see how the above isotopy fails to be a contact isotopy: it is generated by the vector field 
\[
\xi =-\pi x^2\frac{\partial}{\partial x^1}+\pi x^1\frac{\partial}{\partial x^2} 
\]
which does not preserve the contact structure: 
\begin{align*}
&\pounds_\xi \left( \cos \theta dx^1+\sin \theta dx^2\right) \\
&=\xi \rfloor \left( -\sin \theta d\theta \wedge dx^1+\cos \theta d\theta \wedge dx^2\right) +d\left[ \xi \rfloor \left( \cos \theta dx^1+\sin \theta dx^2\right) \right] \\
&=\left( -\pi x^2\sin \theta -\pi x^1\cos \theta \right) d\theta +d\left(-\pi x^2\cos \theta +\pi x^1\sin \theta \right) \\
&=\pi \sin \theta dx^1-\pi \cos \theta dx^2 \\
& \neq \lambda \left( \cos \theta dx^1+\sin \theta dx^2\right).
\end{align*}

Thus Legendrian linking allows us to distinguish between past and future in Minkowski (2+1)-spacetime, whereas smooth linking does not. This hints that it could be the concept of linking we are looking for to express causal relations in the manifold of light rays for $d=3$. 

\begin{Con}
Let $(M,g)$ be a globally hyperbolic (3+1)-spacetime with Cauchy surface diffeomorphic to a subset of $\Bbb{R}^3$, and let $N$ be its manifold of light rays. Then two spacetime points are causally related in $M$ \emph{iff} their skies either intersect or are Legendrian linked in $N$.
\end{Con}

This is a natural extension of Low's conjecture for (2+1)-dimensional spacetimes, and appears to be true for at least some cases. However, a proof appears to require new methods of contact topology.
%
%
%
\section*{Acknowledgments}

We would like to thank Prof. Sir Roger Penrose for suggesting this problem.


\begin{thebibliography}{Low90b}

\bibitem[Arn94]{Ar94}
V.I. Arnold, \emph{Topological invariants of plane curves and caustics},
  American Mathematical Society, 1994.

\bibitem[Arn97]{Ar97}
\bysame, \emph{Mathematical methods of classical mechanics}, Springer, 1997.

\bibitem[CG97]{CG97}
S.~Chmutov and V.~Goryunov, \emph{Polynomial invariants of legendrian links and
  plane fronts}, Amer. Math. Soc. Transl. Ser. 2 \textbf{180} (1997), 25--43.

\bibitem[EF98]{EF98}
Y.~Eliashberg and M.~Fraser, \emph{Classification of topologically trivial
  legendrian knots}, Geometry, topology and dynamics, CRM Proc. Lecture Notes,
  no.~15, Amer. Math. Soc., 1998, pp.~17--51.

\bibitem[EN00]{EN00}
J.~Ehlers and E.~Newman, \emph{The theory of caustics and wavefront
  singularities with physical applications}, J. Math. Phys. \textbf{41} (2000),
  3344--3378.

\bibitem[Low88]{L88}
R.~Low, \emph{Causal relations and spaces of null geodesics}, Ph.D. thesis,
  Oxford University, 1988.

\bibitem[Low89]{L89}
\bysame, \emph{The geometry of the space of null geodesics}, J. Math. Phys.
  \textbf{30} (1989), no.~4, 809--811.

\bibitem[Low90a]{L90a}
\bysame, \emph{The causal geometry of twistor space}, J. Math. Physics
  \textbf{31} (1990), no.~4, 863--867.

\bibitem[Low90b]{L90b}
\bysame, \emph{Twistor linking and causal relations}, Class. Quantum Grav.
  \textbf{7} (1990), no.~2, 177--187.

\bibitem[Low94]{L94}
\bysame, \emph{Twistor linking and causal relations in exterior schwarzschild
  space}, Class. Quantum Grav. \textbf{11} (1994), no.~2, 453--456.

\bibitem[Low98]{L98}
\bysame, \emph{Stable singularities of wavefronts in general relativity}, J.
  Math. Physics \textbf{39} (1998), no.~6, 3332--3335.

\bibitem[Nat00]{N00}
J.~Nat\'ario, \emph{Causal relations in the manifold of light rays}, Ph.D.
  thesis, Oxford University, 2000.

\bibitem[Nat02]{N02}
\bysame, \emph{Linking and causality in (2+1)-dimensional static spacetimes},
  Class. Quantum Grav. \textbf{19} (2002), 3115--3126.

\bibitem[PR86]{PR86}
R.~Penrose and W.~Rindler, \emph{Spinors and spacetime}, Cambridge University
  Press, 1986.

\bibitem[PS91]{PS91}
V.~Prasolov and A.~Sossinsky, \emph{Knots, links, braids and 3-manifolds},
  American Mathematical Society, 1991.

\bibitem[Rol90]{R90}
D.~Rolfsen, \emph{Knots and links}, Publish or Perish, 1990.

\bibitem[Tra97]{T97}
L.~Traynor, \emph{Legendrian circular helix links}, Math. Proc. Cambridge Phil.
  Soc. \textbf{122} (1997), 301--314.

\end{thebibliography}
\end{document}